# Stability of Drift Waves in a Field Reversed Configuration

A. Surjalal Sharma
University of Maryland
Department of Astronomy
College Park, MD  20742

**Abstract**

The drift waves in field-reversed configurations without a toroidal magnetic field, therefore no shear, play an important role in plasma transport.  The short connection length of the poloidal field in these systems leads to significant stabilization by influencing the wave particle resonance.  The field reversed configuration is modeled by the cylindrical Bennett pinch in the limit of large aspect ratio.  The radial eigenmode equation for the universal mode is derived from kinetic theory and the method of quadratic forms is used to study its stability.  The short connection lengths of the field lines lead to ion Landau damping on the inside of the plasma and the stability of the mode depend on the value of the temperature ratio $T_i/T_e$.

# 1. Introduction

Plasma confinement configurations based on the concept of field reversal, produced by internal plasma currents, have many favorable features from the reactor standpoint [Condit et al., 1976]. Examples of such systems are: (a) the Astron [Christofilos, 1958] and ion ring [Sudan and Ott, 1974; Fleischmann and Kammash, 1975; Weibel, 1977] where the main azimuthal current is established by energetic electrons or ions, (b) the field reversed mirror [Carlson, et al., 1978] where the plasma currents are for the most part diamagnetic, and (c) field reversed configurations with rotating magnetic field current drive [Slough and Miller, 2000; Hoffman, 2000; Landsman et al. 2006]. The stability of drift waves in these systems is important for the plasma transport and confinement. Studies of the low-frequency magnetohydrodynamic stability [Sudan and Rosenbluth, 1976, 1979; Lovelace, 1975, 1976; Finn and Sudan, 1978, 1979] and the high frequency microinstabilities [Gerver and Sudan, 1979; Uhm and Davidson, 1979] have been discussed for simple geometrical configurations, e.g., the long thin cylindrical P-layer or the bicycle tire shaped ring of large aspect ratio [Finn and Sudan, 1982]. In this paper we examine the stability of field-reversed rings against drift mode instabilities.

The essential feature of these systems, in the absence of any toroidal field and therefore with no magnetic shear, is the short connection length of the poloidal field lines, which has a strong stabilizing effect on many types of perturbations. In order to determine this effect in quantitative terms we adopt the following simplified model to represent a field-reversed ring. We neglect effects due to toroidal curvature and approximate the ring by a straight cylinder in which the podoidal (azimuthal) field is generated by an axial current. Although it is possible to pursue the analysis for a cylinder of arbitrary cross-section in terms of flux coordinates we shall consider here a circular cross-section. A self-consistent equilibrium for this geometry is developed in section 2, along the lines of a Bennett pinch. The radial eigenmode equation describing the drift perturbation of this equilibrium is derived in section 3 using plasma kinetic theory. The collisionless electrostatic drift mode, viz. the universal mode, is analyzed in section 4. The method of quadratic forms [Antonsen, 1978] is used to study the stability of the mode. A summary and discussion of the analysis is presented in the last section.

**2. Field Reversed Equilibrium**

We consider an ion ring with confined plasma in the background. In the limit of large aspect ratio the ring is approximated by a straight cylinder in which the axial current produces an azimuthal magnetic field with circular field lines around the O-point on the axis. The background plasma is immersed in this magnetic field and has diamagnetic currents in the axial direction. The plasma current is however small compared to the ring current and is ignored. The equilibrium of the ring, which is now approximated by a beam, is axially and azimuthally symmetric and can be described in terms of the equilibrium energy $H_o = mv^2/2$ and the axial momentum $P_z = mv_z + qA_{oz}(r)/c$, where $A_{oz}(r)$ is the z-component of the equilibrium vector potential $\mathbf{A} = (0, 0, A_{oz}(r))$. Assuming a Gibbs-type distribution function for the ring particles,

$$f_{rj}(H_{oj}, P_{zj}) = \frac{N_r}{\left(2\pi T_{rj}/m_j\right)^{1/2}} \exp\left[-\frac{H_{oj} - V_j P_{zj}}{T_{rj}}\right],$$

where $V_j$ is the average velocity and $T_{rj}$ is the temperature, the equilibrium can be described in terms of $A_{oz}(r)$. The solution of this equation in cylindrical geometry yields the Bennett relations [Bennett, 1934]:

$$A_{oz}(r) = -\kappa^{-1} B_o \ln\left(1 + \frac{1}{4}\kappa^2 r^2\right)^2,$$

$$B_\theta(r) = B_o \frac{\kappa r}{\left(1 + \frac{1}{4}\kappa^2 r\right)}, \quad \underset{\sim}{B} = (O, B_\theta, O), \qquad (1)$$

$$n_r(r) = \frac{n_{ro}}{\left(1 + \frac{1}{4}\kappa^2 r^2\right)^2},$$

where $\kappa^{-1}$ is the radial scale length of the ring (beam) with $\kappa^2 = (T_{ri} + T_{re})2\pi e^2 n_{ro}|V_e|/c^2 T_{re}^2$ and $B_o = cT_{re}\kappa/e|V_e|$ is the maximum of the magnetic field, which occurs at $\kappa r = 2$. We consider a plasma of electrons and singly charged ions so that $q_i = e$, $q_e = -e$, and charge neutrality at equilibrium yields the condition, $V_i = -V_e T_{ri}/T_{re}$.

The plasma electrons and ions are assumed to have rigid rotor distributions and in the present large aspect ratio limit they also become rigid drift distributions and may be written as

$$f_{oj} = \frac{N_p}{(2\pi T_j/m_j)^{3/2}} \exp\left[-\frac{m_j v^2}{2} + \frac{1}{L}\left(\frac{v_z}{\Omega_j} + \frac{A_{oz}}{B_o}\right)\right], \qquad (2)$$

where $\Omega_j = q_j B_o/m_j c$ is the gyro frequency and L is the scalelength of the density and $f_{oj}$ has a drift $v_{dj} = T_j/m_j \Omega_j L$. We take $\kappa L \gg 1$ so that the ring particles are localized close to the O-point of the magnetic field and the plasma density falls off on a longer scale length.

Equations (1) and (2) yield the plasma density profile

$$n_p(r) = n_i = n_e = \frac{n_o}{\left(1 + \frac{1}{4}\frac{r^2}{L^2}\right)^{2/\kappa_L}} \qquad (3)$$

and the diamagnetic drift, in the axial direction, is

$$V_{Dj} = -\frac{cT_j}{q_j B_\theta} \frac{1}{n_p} \frac{dn_p}{dr} = \frac{cT_j}{q_j B_o L}. \qquad (4)$$

It may be noted that $V_{Dj}$ is independent of r and equals to $v_{dj}$ obtained from the distribution $f_{pj}$ defined by equation (2). The diamagnetic drift of the electrons is negative and that of the ions positive, thus defining $V_{De} = -V_D$, we get, from the condition for charge neutrality at equilibrium, $V_{Di} = V_D/\tau$, with $\tau = T_e/T_i$.

## 3. Radial Eigenmode Equation

The radial eigenmode equation for drift waves can be derived following a perturbation technique [Pfirsch, 1962] that makes use of the constants of motion and is appropriate for treating the perturbations on self-consistent equilibria constructed from these constants of motion. The perturbed distribution function may be written, suppressing the particle subscript j for simplicity, as

$$f = f_o(H, P_z) + g, \qquad (5)$$

where the Hamiltonian H is defined by $H = H_o + H_1$, with $H_1 = q(\phi - \mathbf{v}\cdot\mathbf{A}/c)$, $\phi$ and $\mathbf{A}$ being the perturbed scalar and vector potentials. In (5) $f_o(H,P_z)$ is that part of *f* which preserves the functional form of the equilibrium distribution function, i.e., it is the adiabatically perturbed part of the distribution function; and g is the departure from the equilibrium form, i.e., the non-adiabatic part.

The equation of motion is

$$\frac{\partial f}{\partial t} + [H, f] = 0,  \tag{6}$$

where the square bracket is the Poisson bracket. On linearizing equation (6) and using the fact that H and $P_z$ are constants of motion, we get

$$\frac{\partial g}{\partial t} + [H_o, g] = -\frac{\partial H_1}{\partial t}\frac{\partial f_o(H_o, P_{oz})}{\partial H_o} + \frac{\partial H_1}{\partial z}\frac{\partial f_o(H_o, P_{oz})}{\partial P_{oz}}.  \tag{7}$$

Integrating equation (7) over the unperturbed orbits yields

$$g = -\frac{\partial f_o(H_o, P_{oz})}{\partial H_o}\int_{-\infty}^{t} dt' \frac{\partial H_1(t')}{\partial t'} + \frac{\partial f_o(H_o, P_{oz})}{\partial P_{oz}}\int_{-\infty}^{t} dt' \frac{\partial H_1(t')}{\partial z'}.$$

For perturbations of frequency ω, axial wavenumber k and azimuthal mode number m,

$$H_1 = q\widetilde{H}_1 \exp(-i\omega t + ik_z + im\theta),  \tag{8}$$

the above expression becomes, using equation (2) for $f_o$,

$$g_i = -\frac{iq_j(\omega - kV_{Dj})}{T_j} f_{oj}(H_o, P_{oz}) \int_{-\infty}^{t} dt' \widetilde{H}_1 \exp(-i\omega t' + ikz' + im\theta'),$$

(9)

where the particle subscript j has been restored.

The perturbed particle and current densities then are [Pfirsch, 1962],

$$\begin{aligned} n_{ij} &= \lim_{\phi_o \to 0}\left(\phi\frac{\partial}{\partial \phi_o} + A_z \frac{\partial}{\partial A_{oz}}\right) n_{pj}(\phi_o, A_{oz}) + \int d^3v g_j, \\ \mathbf{j}_{1j} &= \lim_{\phi_o \to 0}\left(\phi\frac{\partial}{\partial \phi_o} + A_z \frac{\partial}{\partial A_{oz}}\right) \mathbf{j}(\phi_o, A_{oz}) + \int d^3v \mathbf{v} g_j, \end{aligned}  \tag{10}$$

where $n_{pj}(\phi_o, A_{oz}) = \int d^3v f_{oj}(H_o, P_{oz})$ and $\mathbf{j}_{oj} = q_j \int d^3v \mathbf{v} f_{oj}(H_o, P_{oz})$

Equations (9) and (10), along with the Maxwell's equations and the details of the unperturbed particle orbits provide a complete kinetic description of the perturbations of a cylindrical plasma in an azimuthal magnetic field.

We shall deal here with the collisionless electrostatic drift mode, viz. the universal mode. For perturbations with the scale length along the r-axis much larger than the extent of the particle orbit, we expand $\phi(r')$ about $r' = r$ as

$$\phi(r') \simeq \phi(r) + (r'-r)\nabla_r \phi(r) + \frac{1}{2}(r'-r)^2 \nabla_r^2 \phi(r) \qquad (11)$$

The perturbed density as given by equation (10) now becomes

$$n_{1j} = -\frac{q_j n_p}{T_j} \exp(-i\omega t + ikz + im\theta)$$
$$\times \left\{ \phi(r) + i(\omega - kV_{Dj})\left[ X_o \phi(r) + X_1 \nabla_r \phi(r) + \frac{1}{2} X_2 \nabla_r^2 \phi(r) \right] \right\}, \qquad (12)$$

with

$$X_\ell = \int d^3v \tilde{f}_{oj} \int_{-\infty}^{t} dt'(r'-r)^\ell \exp[i\psi(t')] \qquad \ell = 0, 1 \text{ and } 2,$$

where $f_{oj}(H_o, P_{oz}) = n_p \tilde{f}_{oj}$ and $\psi(t') = -(t'-t) + k(z'-z) + m(\theta'-\theta)$. The orbit integrals, the $X_\ell$'s, may now be evaluated using the appropriate particle orbits, viz. the gyro orbits:

$$z' = z - \frac{v_\perp}{\Omega_j}[\cos\alpha - \cos(\alpha - \Omega_j(t'-t))]$$
$$r' = r + \frac{v_\perp}{\Omega_j}[\sin\alpha - \sin(\alpha - \Omega_j(t'-t))] \qquad (13)$$
$$\theta' = \theta + v_\parallel(t'-t)/r,$$

where α is the initial phase angle. The drifts due to the curvature and gradient of the magnetic field are not considered here. With these orbits, we now get,

$$X_o = \frac{\Gamma_o(b_j)}{\omega} \zeta_j Z(\zeta_j) \qquad X_1 = 0(\omega/\Omega_j)$$
$$\text{and} \qquad X_2 = -\frac{2a_j^2}{i\omega} \frac{d\Gamma_o}{db_j} \zeta_j Z(\zeta_j) \tag{14}$$

where $Z(\zeta_j)$ is the plasma dispersion function with $\zeta_j = \omega/\sqrt{2} k_\| v_j$, $k_\| = m/r$, the parallel wavenumber and $v_j = (T_j/m_j)^{1/2}$ the particle thermal velocity. Also $\Gamma_o(b_j) = I_o(b_j)\exp(-b_j)$, $I_o$ being the Bessell function of second kind and order zero, $b_j = k^2 a_j^2$ with the gyro radius $a_j = v_j/\Omega_j$.

For the low frequency drift modes of interest here, $X_1$ and the terms with the electron gyroradius squared are neglected. From equation (12) we then obtain the expressions for $n_{1j}$ and the charge neutrality condition, $n_{1e} = n_{1i}$, yields the radial eigenmode equation

$$\nabla_r^2 \phi - q(r)\phi = 0, \qquad \nabla_r^2 \equiv \frac{1}{r}\frac{d}{dr} r \frac{d}{dr}, \tag{15}$$

where

$$q(r) = \frac{1 + \tau + \frac{\omega - \omega_{*e}}{\omega}\zeta_e Z(\zeta_e) + \frac{\tau(\omega - \omega_{*i})}{\omega}\zeta_i Z(\zeta_i)\Gamma_o(b_i)}{a_i^2 \frac{d\Gamma_o}{db_i}\frac{\tau(\omega - \omega_{*i})}{\omega}\zeta_i Z(\zeta_i)} \tag{16}$$

$\omega_{*j} = kV_{Dj}$. The treatment here is fully kinetic and the two Z-functions represent the interactions of the electrons and ions with the waves. It may be noted that the resonant as well as the non-resonant interactions of both the particle species are retained in equations (15) and (16). There are two radial scale lengths associated with these effects. The radial distance at which the ion thermal velocity equals the wave phase velocity, viz. $\zeta_i = 1$ defines the ion resonance point $r_i$, i.e., $r_i = \sqrt{2}\, mv_i/\omega$. Similarly the electron resonance point is $r_e = \sqrt{2}\, mv_e/\omega$. The different radial scale lengths then compare as

$$a_e \ll a_i \ll r_i \ll r_e \ll K^{-1} \ll L. \tag{17}$$

Thus for $r < r_i$, i.e., close to the axis of the cylinder, ion Landau damping is the dominant wave-particle effect. For $r_i < r < r_e$ the ions are non-resonant and the electron inverse Landau damping is important.

In the kinetic description of drift waves in a sheared magnetic field in slab geometry [Krall ad Rosenbluth 1965] the locations of the ion and electron resonance points are reversed from the one given by equation (17). The electrons then contribute to the inverse Landau damping near the mode rational surface and the ion sound point $r_i$ is away from it. Also the radial eigenmode equation in that case is valid only in the neighbourhood of a mode rational surface, whereas equations (15) and (16) are for the whole radial extent of the plasma.

The radial eigenmode equation, (15) and (16), was derived by taking the particles to be executing gyro orbits. The magnetic field however goes to zero on the axis and increases linearly with r for small r, as given by equation (1). Consequently, the particles located close to the axis execute radial betatron oscillations. The fraction of background plasma particles in this region is however small and are expected to have little effect on the stability of drift waves in the background plasma. When the interaction of the ring with the drift mode is considered, the betatron oscillations could have important consequences as in the case of magnetohydrodynamic modes [Finn and Sudan, 1978, 1979 and 1982].

## 4. Stability of Bound Eigenmodes of the Universal Mode

The radial eigenmode equation for the universal mode, equations (15) and (16), governs its stability. Here we shall study the stability of the mode by constructing appropriate quadratic forms. This procedure [Antonsen, 1978] has shown that unstable and bound eigenmodes of the universal mode do not exist in sheared magnetic field in a slab geometry. The starting point of this method is the recognition that the radial eigenmode equation may be analytically continued to a line, in the complex r plane, on which the Z-functions reduce to a purely imaginary quantity with a specified sign. On this line quadratic forms may be constructed and the stability of the mode studied. For bounded eigenmodes the analytic continuation from the real r axis to this line should not cross the dominant asymptotic Stoke's line. Asymtotically, equations (15) and (16) become

$$\frac{d^2\phi}{dr^2} - \beta^2 \phi = 0, \quad \beta^2 = \frac{\Gamma_o(b_i) - 1}{d\Gamma_o / db_i} > 0,$$

so that

$$\phi \sim \exp(\pm \beta r) \tag{18}$$

and we choose the decaying solution for $0 < r < \infty$ as usual. From equation (18) it is evident that the asymptotic Stoke's lines are along the real r axis in the complex r plane, the sub-dominant branch being $0 < r < \infty$.

To apply the method of quadratic forms to equations (15) and (16), we assume an unstable eigenvalue, i.e., $\omega = \omega_r + i\gamma \; (\gamma > 0)$, analytically continue the equation into the complex r plane and define the line

$$i\eta = \zeta_i = \frac{\omega}{\sqrt{2} \, k_\parallel v_i}, \quad k_\parallel = \frac{m}{r}, \tag{19}$$

where η is real and positive. Then

$$\zeta_i Z(\zeta_i) = -\sqrt{\pi}\, \eta \exp(\eta^2) erf(\eta) = -f(\eta)$$
$$\zeta_e Z(\zeta_e) = -\sqrt{\pi}\, \eta\, xp(\varepsilon^2\eta^2) erf(\varepsilon\eta) = -f(\varepsilon\eta) \quad (20)$$

where, both f(η) and f(εη) are real and positive and erfc(η) is the complimentary error function. From equation (19), the location of the η line in the complex r plane is given by the relation

$$\frac{\operatorname{Re} r}{\operatorname{Im} r} = \frac{\gamma}{\omega_r}. \quad (21)$$

Since γ > 0, the η line is in the first quadrant for $\omega_r$ > 0 and in the fourth quadrant for $\omega_r$ < 0. Evidently in both the cases the analytic continuation of equations (15) and (16) from the positive real r axis to the line does not cross the dominant Stoke's line (negative real r axis) and the anti-Stoke's lines (imaginary r axis). The boundary condition on the η line are:

$$either \quad \phi = 0, \quad or \quad \frac{d\phi}{d\eta} = 0 \quad at \quad \eta = 0;\quad and \quad \phi = 0 \quad at \quad \eta = \infty. \quad (22)$$

On the η line, the asymptotic φ given by equation (18) is

$\phi \sim \exp[-\sqrt{2}\, mv_i](i\omega_r + \gamma)\eta/|\omega^2|$, which decays with η for γ > 0. This shows that the analytic continuation has not crossed any anti-Stoke's line. Now the quadratic form may be constructed to study the stability of bounded eigenmodes for the following cases defined by the ratio of the electron and ion temperatures.

*A. Small $T_e/T_i$: Ions hotter than electrons*

When the background plasma ion temperature is much larger than the electron temperature, the electron Z-function may be expanded for large argument. On taking the limit of small $T_e$, the function q(r) defined by (16) becomes

$$q(r) = \frac{\frac{k_\parallel^2 v_s^2}{\omega^2} - \left(1 - \frac{\omega_{*i}}{\omega}\right)[1 + \zeta_i L(\zeta_i)\mu_o(b_i)] - k^2 a^2}{\tilde{a}_i^2 \left(1 - \frac{\omega_{*i}}{\omega}\right)\zeta_i L(\zeta_i)} \qquad (23)$$

where $v_s^2 = T_i/m_e$, $a^2 = m_e a_i^2/m_i$ and $\tilde{a}_i^2 = -a_i^2 = -a_i^2 d\Gamma_o/db_i$. The acoustic term in equation (23), viz. $k_\parallel^2 v_s^2/\omega^2$ corresponds to the electron-acoustic mode [Arefev, 1970; Sharma et al., 1983], which is the acoustic mode of a plasma in a magnetic field with $T_i \gg T_e$.

On the η line equation. (15) and (23) can be rewritten as

$$\frac{\omega(\omega - \omega_{*i})}{2m^2 v_i^2}\tilde{a}_i^2 \frac{1}{\eta}\frac{d}{d\eta}\eta\frac{d\phi}{d\eta} + \frac{1}{f(\eta)}\left[\frac{m_i}{2m_e \eta^2} + k^2 a^2 + \left(1 - \frac{\omega_{*i}}{\omega}\right)G(\eta)\right]\phi = 0 \qquad (24)$$

where

$$\begin{aligned}G(\eta) &= 1 + \zeta_i Z(\zeta_i)\Gamma_o(b_i) = 1 - f(\eta)\Gamma_o(b_i) \\ &= g(\eta) + (1 - \Gamma_o)f(\eta), \quad g(\eta) = 1 - f(\eta)\end{aligned} \qquad (25)$$

is real and positive. Now we construct quadratic forms by multiplying equation (24) by $\eta\phi*$ and integrating over η from η = 0 to η = ∞, and using the boundary conditions defined by equation (22). The resulting equation can be separated into the real and imaginary parts to yield the two equations:

$$\int_o^\infty d\eta\eta \left[ -\frac{\left(\omega_r^2-\gamma^2\right)}{2m^2v_i^2}\tilde{a}_i^2\left|\frac{d\phi}{d\eta}\right|^2 \right.$$
$$\left. +\frac{1}{f(\eta)}\left\{\frac{\omega_r^2-\omega_{*i}\omega_r+\gamma^2}{|\omega-\omega_{*i}|^2}\left(\frac{m_i}{2m_e\eta^2}\right)+G(\eta)\right\}|\phi|^2 \right]=0. \quad (26)$$

and

$$\gamma\int_o^\infty d\eta\eta\left[\frac{\omega_r}{m^2v_i^2}\tilde{a}_i^2\left|\frac{d\phi}{d\eta}\right|^2+\frac{1}{f(\eta)}\frac{\omega_{*i}}{|\omega-\omega_{*i}|^2}\left(\frac{m_i}{2m_e\eta^2}+k^2a^2\right)|\phi|^2\right]=0. \quad (27)$$

Evidently for $\omega_r/\omega_{*i}>0$, equation (27) cannot be satisfied and for $\omega_r/\omega_{*i}<0$ we use (27) to reduce (26) into

$$\int_o^\infty d\eta\eta\left[\frac{\omega_r^2+\gamma^2}{2m^2v_i^2}\tilde{a}_i^2\left|\frac{d\phi}{d\eta}\right|^2+\frac{1}{f(\eta)}\left\{\frac{\omega_r^2+\gamma^2}{|\omega-\omega_{*i}|^2}\left(\frac{m_i}{2m_e\eta^2}+k^2a^2\right)+G(\eta)\right\}|\phi|^2\right]=0 \quad (28)$$

This equation cannot be satisfied for $\omega_r/\omega_{*i}<0$. Thus our earlier assumption that unstable bound eigenmodes exist, cannot hold and hence such eigenmodes do not exist. In this case of the ions being much hotter than the electrons, the ion Landau damping is strong and the electron inverse Landau damping, which usually destabilizes the universal mode, is weak. Consequently the mode is stable.

B. Large $T_e/T_i$: Electrons hotter than ions

In most plasmas the electron temperature is higher than the ion temperature. In this case the ion Z-function may be expanded for large argument and in the limit of $T_e \gg T_i$, q(r) becomes

$$q(r)=\frac{1}{a_s^2}\left\{\frac{k_\parallel^2C_s^2}{\omega^2}-\left(1-\frac{\omega_{*e}}{\omega}\right)[1+\zeta_e L(\zeta_e)]-k^2a_s^2\right\}, \quad (29)$$

where $C_s^2 = T_e / m_i$ and $a_s = C_s / \Omega_i$. The eigenmode equation, (15) with q(r) given by equation (29) is the same as the corresponding equation for sheared magnetic field [Pearlstein and Berk, 1969] except for the definition of $k_\parallel$.

Equations (15) and (29) on the $\eta$ line yield

$$\frac{\omega^2 m_e}{2m^2 m_i} \frac{1}{\eta} \frac{d}{d\eta} \eta \frac{d\phi}{d\eta} + \left[\frac{m_e}{2m_i \eta^2} + \left(1 - \frac{\omega_{*e}}{\omega}\right) g(\varepsilon\eta) + k^2 a_s^2\right]\phi = 0, \tag{30}$$

where

$$g(\varepsilon\eta) = 1 + \zeta_e Z(\zeta_e) = 1 - f(\varepsilon\eta), \quad \varepsilon = (m_e / m_i \tau)^{1/2} \tag{31}$$

is real and positive. As before we construct quadratic forms and obtain the two equations:

$$\int_o^\infty d\eta\,\eta \left[-\frac{(\omega_r^2 - \gamma^2) m_e}{2m^2 m_i}\left|\frac{d\phi}{d\eta}\right|^2 + \left\{\frac{m_e}{2m_i \eta^2} + \left(1 - \frac{\omega_r \omega_{*e}}{\omega^2}\right) g(\varepsilon\eta) + k^2 a_s^2\right\}|\phi|^2\right] = 0. \tag{32}$$

and

$$\gamma \int_o^\infty d\eta\,\eta \left[-\frac{\omega_r m_e}{m^2 m_i}\left|\frac{d\phi}{d\eta}\right|^2 + \frac{\omega_{*e}}{|\omega|^2} g(\varepsilon\eta)|\phi|^2\right] = 0. \tag{33}$$

For $\omega_r / \omega_{*i} < 0$, equation (33) cannot be satisfied, and for $\omega_r / \omega_{*i} > 0$, equations (32) and (33) may be combined to yield

$$\int_o^\infty d\eta\,\eta \left[-\frac{(\gamma^2 - 3\omega_r^2) m_e}{2m^2 m_i}\left|\frac{d\phi}{d\eta}\right|^2 + \left\{\frac{m_e}{2m^2 m_i} + g(\varepsilon\eta) + k^2 a_s^2\right\}|\phi|^2\right] = 0. \tag{34}$$

From this equation no general conclusion about the $\omega_r / \omega_{*e} > 0$ modes could be reached. The drift mode could be unstable in this case, being destabilized by the strong inverse Landau damping by the electrons and weak damping by the ions. In the case of sheared $B$ in slab geometry, the non-existence of unstable bound eigenmodes was shown by constructing quadratic forms [Antonsen, 1978] and also by other methods [Ross and Mahajan, 1978; Tsang, et al. 1978; Chen, et al., 1978].

*C. Arbitrary $T_e/T_i$*

When the electron and ion temperature are not restricted, equations (15) and (16) on the η line defined by equation (19) is

$$-\frac{\omega^2 \tilde{a}_i^2}{2m^2 v_i^2} \frac{1}{\eta} \frac{d}{d\eta} \eta \frac{d\phi}{d\eta} - \frac{1}{f(\eta)} \{(\tau\omega_r^2 + \omega_{*e}\omega_r + \tau\gamma^2)F(\eta) + \omega_{*e}(\tau\omega_r + \omega_{*e})(f(\varepsilon\eta) - \Gamma_o f(\eta)) + i\gamma(1+\tau)\omega_{*e} g(\varepsilon\eta)\}\phi = 0, \quad (35)$$

where

$$F(\eta) = \tau(1-\Gamma_o) + \tau\Gamma_o g(\eta) + g(\varepsilon\eta) \quad (36)$$

Here the functions f and g are as defined by equations (20), (25) and (31), and $F(\eta)$ is real and always positive. The quadratic forms may be constructed as before and equation (35) then separates into the real and imaginary parts:

$$\int_o^\infty d\eta \eta [\frac{\omega_r^2 - \gamma^2}{2m^2 v_i^2} \tilde{a}_i^2 \left|\frac{d\phi}{d\eta}\right|^2 \\ - \{(\tau\omega_r^2 + \omega_{*e}\omega_r + \tau\gamma^2)F(\eta) + \omega_{*e}(\tau\omega_r + \omega_{*e})(f(\varepsilon\eta) - \Gamma_o f(\eta))\}\frac{|\phi|^2}{f(\eta)}] = 0 \quad (37)$$

and

$$\int_o^\infty d\eta \eta \left[\frac{\omega_r \tilde{a}_i^2}{m^2 v_i^2} \left|\frac{d\phi}{d\eta}\right|^2 - (1+\tau)\omega_{*e} g(\varepsilon\eta)\frac{|\phi|^2}{f(\eta)}\right] = 0. \quad (38)$$

Equation (38) cannot be satisfied for $\omega_r/\omega_{*e} < 0$ and $\omega_r/\omega_{*e} > 0$, and it may be combined with equation (37) to yield, after some algebra,

$$\int_o^\infty d\eta \left[ \frac{\omega_r^2 + \gamma^2}{2m^2 v_i^2} \widetilde{a}_i^2 \left|\frac{d\phi}{d\eta}\right|^2 + \{(\omega_r^2 + \gamma^2)F(\eta) \right.$$
$$\left. + \omega_{*e}(2\tau\omega_r + \omega_{*e})h(\eta)\} \frac{|\phi|^2}{f(\eta)} \right] = 0, \qquad (39)$$

where

$$h(\eta) = f(\varepsilon\eta) - \Gamma_o f(\eta) = (1 - \Gamma_o)f(\varepsilon\eta) - f(\eta) \qquad (40)$$

For h(η) ≥ 0, equation (39) cannot be satisfied when $\omega_r / \omega_{*e} > 0$ and we conclude that in this case unstable bound eigenmodes do not exist.

From equation(40) it is seen that the sufficient condition for h(η) ≥ 0 is $f(\varepsilon\eta) \geq f(\mu)$. Since $f(\varepsilon\eta)$ and $f(\eta)$ as defined by equation (20) have the same functional form and are monotonically increasing with η, $f(\varepsilon\eta) \geq f(\eta)$ for ε ≥ 1, i.e., $\tau = T_e/T_i \leq m_e/m_i$. This is in agreement with the conclusion we have arrived earlier in sub-section A. It should be emphasized, however, that this condition is only a sufficient condition and may not imply complete stability. However, for τ > $m_e/m_i$ there are indications that the mode would not be unstable. First, $\Gamma_o(b_i) < 1$ and this will relax the condition ε ≥ 1 even when $f(\varepsilon\eta)$ is not greater than $f(\eta)$, and $h(\eta)$ can be positive or zero. Second, when $h(\eta)$ is negative, the co-efficient of $|\phi|^2$ in equation (39) can be positive or zero. Further the integrand of equation (39) could still be positive when the coefficient of $|\phi|^2$ is negative. In view of these considerations the non-existence of unstable bound eigenmode could be expected for $\tau = T_e/T_i \leq m_e/m_i$, but less than some $\tau_c$. However the present analysis does not yield this upper limit $\tau_c$.

## 5. Conclusion

We have studied the stability of the universal mode in a field-reversed configuration in which a small population of fast particles produces the closed field lines. The equilibrium magnetic field has no shear and so the shear damping is absent. However, because of the short connection lengths of the closed field lines the ion Landau damping occurs close to the axis and has a stabilizing influence. We have shown that the sufficient condition for the nonexistence of unstable bound eigenmodes is $\tau = T_e / T_i \leq m_e / m_i$. This limit on $\tau$ is rather severe but in most cases it will be relaxed to a value $\tau_c$, where the ion Landau damping balances the electron inverse Landau damping. The value of $\tau_c$ can be established by numerically solving the radial eigenmode equation.

The method of quadratic forms used here is an elegant method to study the stability of modes governed by radial eigenmode equations. However it may not yield conclusive results in some situations. For example, when the drifts due to the gradient and curvature of B are included, the Z-functions of the electrons and ions cannot be simultaneously reduced to purely imaginary quantities on the same $\eta$ line, unlike the case dealt in section 4-C. However, when the electron Z-function is expanded for small argument, i.e., the destabilizing electron inverse Landau damping is retained, the non-existence of bound unstable eigenmodes can be proved under certain assumptions [Sharma and Sudan, 1979].

The case of plasmas with ions much hotter than electrons may not be readily realized in the laboratory. However, in space plasmas, e.g. Earth's magnetotail [Baumjohann et al., 1989; Sitnov et al., 1998] the ions have much higher temperature than the electrons and in these situations the drift waves are expected to be stable.